\documentstyle[11pt,aaspp]{article}
\lefthead{Drukier, G.A.}
\righthead{Models of NGC 6397 -- B. }
\slugcomment{}

\def\Msolar{\ifmmode M_{\mathord\odot}\else$M_{\mathord\odot}$\fi}
\def\et{et\thinspace al.\ }
\def\chiMF{\ifmmode \chi^2_{MF}\else$\chi^2_{MF}$\fi}
\def\chiSDP{\ifmmode \chi^2_{SDP}\else$\chi^2_{SDP}$\fi}

\def\rlrt{\ifmmode r_l/r_t\else$r_l/r_t$\fi}

\begin{document}

\title{Fokker-Planck models of NGC 6397 -- B. The globular cluster
\footnote{Postscript figures for this paper are available by anonymous
FTP from ftp.ast.cam.ac.uk in the directory
/pub/drukier or by email to the author at
drukier@mail.ast.cam.ac.uk. This paper has been submitted for publication
in {\it The Astrophysical Journal}.}}
\author{G. A. Drukier}
\affil{Institute of Astronomy, University of Cambridge}
\authoraddr{Institute of Astronomy, University of Cambridge, Madingley Rd.,
Cambridge, CB3 0HA, England; Internet: drukier@mail.ast.cam.ac.uk}

\begin{abstract}
This is the second of two papers presenting a detailed examination of
Fokker-Planck models for the globular cluster NGC 6397 and is concerned
with extracting information on the cluster from the models.  The models
give a current cluster mass of $6.6\pm 0.5\times 10^4 \Msolar$ of which
about 2500 \Msolar\ is in neutron stars. This mass gives a $V$-band
mass-to-light ratio of 1.2 in solar units.  The models and data provide
weaker estimates of the structural parameters, but suggest that the
core radius is less than 0.3 pc (11\arcsec) and the tidal radius is
$17\pm 4$ pc.  In turn, by assuming a flat rotation curve for the
galaxy, the mass and tidal radius suggest that the latter was set at a
distance of 2.5 kpc from the galactic center.
\end{abstract}

\section{Introduction}
This is the second part of a binary paper discussing  Fokker-Planck
models matched with observations of the globular cluster NGC~6397. In
the first paper (Drukier 1994; Paper A), the details of the modeling
and comparison techniques were discussed, as was an overview of results
of the over 1000 models run.  Briefly, the models solve the isotropic,
orbit-averaged form of the Fokker-Planck equation, where the
distribution functions are functions of energy and mass. An energy
source in the form of a statistical treatment of binaries formed in
three-body reactions is used to reverse core collapse. The models also
include a tidal boundary and the effects of mass-loss due to stellar
evolution. More details are given in Paper A together with definitions
for many of the symbols used here. The data used for the comparisons
are the surface density profile (SDP) and two mass functions (MFs) from
\markcite{Drukier \et (1993)}, the intermediate-distance mass-function
from \markcite{Fahlman~\et (1989)} and the velocity dispersion profile
from \markcite{Meylan \& Mayor (1991)}. Again, I follow the naming
convention of \markcite{Drukier \et (1993)} and refer to the three mass
functions as the du~Pont:if, FRST, and du~Pont:out MFs in order of
distance from the cluster center.

It was found in Paper A that neither the initial mass function  (IMF),
nor any of the other initial parameters were uniquely constrained by
the observations.  As a consequence, there are a number of models which
give satisfactory matches.  In this paper I will describe several of
these models and use them to discuss the stellar content of NGC~6397,
its current mass and mass-to-light ratio, and its dynamical history.
I will also look at the tidal radius of the cluster, which is related
its galactocentric distance and orbit, and the core radius.

The next section discusses information extracted from the entire
ensemble of models, ie. the current mass of the cluster and its core
and tidal radii.  The first is well constrained; the latter two less
so. Section \ref{models} presents comparisons of nine models drawn from
the various model sets. The model sets are defined  as in Paper~A by
their IMF and tidal radius $r_t$. The definitions of the IMFs are given
in Table~\ref{IMFS}. The tidal radius defines the galactocentric radius
at which the model is assumed to orbit and hence the strength of the
tidal stripping. It is quoted for a cluster with mass $10^5 \Msolar$.
The true tidal radius scales as $M^{1/3}$ (see \S 3.2 in Paper A). These
have been chosen as the best overall models from each.  The differences
between the best matches from each IMF give information on the relative
proportions of high-mass remnants, visible stars, and faint low-mass
stars. The final section summarizes the conclusions of this paper and
binds this binary.

\section{Global constraints}
\subsection{Mass}
One of the initial results from the preliminary ``hunting'' stage
described in \S 3.3 of Paper A, was the observation that there was a
preferred mass for matching the mass functions.  Figure.~\ref{mass vs.
chi} shows \chiMF\ (the quality of fit to the mass functions as defined
in \S 3.1 of Paper A) vs. total mass for a sample of 21 U20 models with
varying initial parameters including $r_t$.  In all cases the
\chiMF\ vs.  mass  curve is parabolic and has a minimum  near a mass
of  $7\times 10^4 \Msolar$.  For this IMF, the mass of the model must
be approximately $7\times 10^4 \Msolar$ in order to match the observed
mass functions. If NGC~6397 did have this IMF, then this result implies
that the current mass of the cluster is also $7\times 10^4 \Msolar$. I
fitted the \chiMF\ vs. mass  curves with a parabola in the mass range
$4\times 10^4 \Msolar$ to $1.5 \times 10^5 \Msolar$  in order to refine
the estimate of the optimal mass.  Figure~\ref{mass vs. chi 2} shows
this minimum in \chiMF\ as a function of mass for all the  U20 models
having such minima. The models with extremely poor \chiMF\ tend to have
other problems which make them unacceptable matches in any case.
Usually, they started off with long relaxation times and show  little
dynamical evolution.  As a result, the best fitting mass is a
compromise between all the poor fits.  That an adequate estimate is
possible from even such poor models strengthens my confidence in this
method of estimating the mass of the cluster.  The models show a range
of optimal masses, but the mode, mean and median all indicate a typical
value of $6.9\times 10^4\Msolar$. The apparent correlation between the
minimum value of \chiMF\ and the optimal mass is related to the
correlation between \chiMF\ and the age of the model as discussed in
Paper A.

With the exception of some very poor fitting models, the other model
sets also give a consistent mass for NGC~6397 even though most of the
models do not give very good matches for the  shapes of the individual
mass functions, let alone the surface density profile.  This shows that
we can get a current mass for NGC~6397 without knowing the IMF to any
great precision. This is so, provided that the IMF being used does give
present day MFs consistent with the observations. By using a range of
IMFs, as is done below,  we can increase our confidence in the
results.

In Paper A, I showed that the requirement for a model to have its best
matches to both \chiMF\ and \chiSDP\ (the quality of fit to the surface
density profile as defined in \S 3.1 of Paper A) simultaneously defines
a surface in the $\left(W_0, M_0, \rlrt \right)$ parameter
space.\footnote{$W_0$ is the dimensionless central potential of the King
model used for the initial state of the model, $M_0$ is the initial
mass, and \rlrt\ is the ratio of the initial  limiting radius of the
model to the tidal radius defined above.} I refer to this as the
``$\Delta t=0$ surface'' and the models on it are referred to as being
``well-fitting''.  This surface actually has a thickness of about $\pm
0.4$ Gyr which arises from the interval between records of the state of
the models (See \S 5 in Paper A). By considering the $\Delta t=0$
surfaces for the various model sets, we can approach the mass question
from a slightly different angle.  Figure~\ref{mass grid} shows  the
mass at minimum \chiMF\ as estimated on the $\Delta t=0 $ surface as a
function of time for each of the model sets.  The format of this
diagram is the same as for similar figures in Paper A. For the U20,
$r_t=18.5$ pc model set, the time dependence of the optimal mass is
similar to that seen for the individual models, but with less scatter
in mass.  The remaining scatter derives from the effects of variations
in the other initial parameters. This model set covered the largest
range of the $(W_0,M_0)$ parameter space (recall from Paper A that
\rlrt\ is a function of $(W_0,M_0)$ on the $\Delta t=0$ surface) and
shows the largest variation in the optimal mass at a given time. This
dependence on the other parameters is confirmed by the apparent
bimodality in the U20, $r_t=21$ pc model set.  The models in this set
were run in two separated ``islands'' in $(W_0,M_0)$ space, and these
are reflected in the diagram.  The lower optimal mass models started
with $M_0<5\times 10^5\Msolar$ and $W_0<5.75$ and the higher mass group
originate in an island with $M_0>  5\times 10^5\Msolar$ and
$W_0>6.00$.  The other model sets covered much smaller regions of
$(W_0,M_0)$ space. From the two U20 model sets covering a large range
of parameters (ie.  those with $r_t=18.5$ pc and $r_t=21.$ pc) I
estimate the intrinsic scatter in a mass estimate at a given age to be
$\pm 1,000 \Msolar$.  For all but the U10 model sets, there is an
increase of 800 to 900 \Msolar/Gyr in the optimal mass over the time
range used.  Adopting the isochrone age of 16$\pm$2.5 Gyr
\markcite{(Anthony-Twarog, Twarog, \& Suntzeff 1992)}, this becomes a $\pm 2000
\Msolar$ systematic uncertainty in the optimal mass.
Table~\ref{optimal mass} gives the estimated optimal mass for each of
the model sets for both 16 Gyr and for the age of the best models in
that model set.  The U20, $r_t=17$ pc mass is based on an extrapolation
to 16 Gyr.  For the X2 and NNS model sets, there was not a sufficient
number of useful models to produce such a graph. For these model sets
the masses in Table~\ref{optimal mass} are averages over all the models
showing minima in \chiMF.  In the X2 model set, no systematic trend
with time was seen while for the NNS model set there may be a trend of
the same size as in the majority of the other model sets.  The model sets
which give high mass estimates are those which do not fit the surface
density profile at 16 Gyr, but only at ages closer to 12 Gyr.  At the
earlier time the masses are lower and are more consistent with the U10
masses.  Based on this I conclude that the current mass of NGC~6397 is
$6.6\pm 0.5\times 10^4\Msolar$.

\subsection{Tidal and core radii}
The computation of \chiMF\ involves  detailed comparison with the  observed
mass functions at three radii. We already know that the IMF is
approximately right for the cluster, otherwise the shapes of the mass
functions would give very poor matches. We also know that we have
approximately the correct tidal radius because we can fit the surface
density profile fairly well.  As I have shown, the mass at which
\chiMF\ reaches a minimum  reflects the optimal fitting of the three
mass functions combined. This point is illustrated  by the most
discordant point in Fig.~\ref{mass vs. chi 2}.  This model started with
$r_t=21$pc, $W_0 = 4.$, $M_0 = 6.97\times 10^5\Msolar$ and
$\rlrt=0.66$. It suffered  a large stellar evolution expansion and lost
a lot of mass through the tidal boundary, so it did not  re-collapse
very much before its mass went to zero.  The model never really fit any
of the observed MFs, but at a mass of $7.6\times 10^4\Msolar$ the  fit
to the mass functions was the least poor, giving, even for this very
poor model, a minimum in \chiMF, at $\chiMF=3.6$.  The bulk of the
models have minima in \chiMF\ much closer to unity and match the
observed MFs much better. That all the minima are not at the same point
reflects the detailed differences in the mass function comparisons.

One type of difference depends on the history of the model and the
small effects that changes in the initial conditions have on the degree
of mass segregation.  There can be differences both with radius at a
given time and with respect to the IMF.  The size of these differences
warrants further study, but will not be dealt with here.

The other type of difference is related to the
density profile. If we compare models GM031 and GK079 (Figs.~\ref{GM031}
and \ref{GK079}), which have $r_t = 17$ and 21 pc respectively, it is
obvious that the latter has a higher density at large radii and,
correspondingly, a du Pont:out MF which, unlike most of the rest of the
models, is not systematically smaller than the observed MF.  In
principle we can use this effect to try and extract an estimate of the
current tidal radius of NGC~6397.  Unfortunately, only weak limits can
be applied in the case of NGC~6397.  The primary problem is that the
outermost mass function is still not far enough out to be a strong
constraint on the tidal radius. But, as was discussed in
Drukier~\et~(1993), at the radial distance of the du Pont:out mass
function, only half of the objects counted belong to the cluster.
Fields further from the cluster center will be even more heavily
contaminated with field stars.  Within any set of models there is a
wide range of minima both in the composite \chiMF\ and in $\chi^2$ for
just the du Pont:out MF because of the other factors discussed above.
Until there is some way to disentangle these effects it will be
difficult to draw any conclusion on the tidal radius of the cluster.
{}From an examination by eye of specific models it appears to me that
$r_t\sim 20 \pm 4 $ pc.  Assuming a flat rotation curve at 220 km
s$^{-1}$ for the galaxy, this gives, using eq.~(5) of Paper~A,  a
galactocentric distance of $2.5\pm 0.7$ kpc for NGC~6397.  The observed
galactocentric distance of the cluster is 6 kpc \markcite{(Djorgovski
1993)}.  \markcite{Cudworth \& Hanson (1993)} found its space velocity
to be $(\Pi, \Theta, Z)= (24\pm 6, 126 \pm 12, -105\pm 12)$ km
s$^{-1}$. From the $\Pi$ and $Z$ velocities for this southern cluster
($b= -12\deg$), NGC~6397 is receding from perigalacticon. The present
tidal radius ($17\pm 4$ pc) may reflect the distance of its most recent
perigalactic passage.

For almost all the models run the optimal fits to the surface density
profile occurred in the collapse phase. The change in \chiSDP\ as a
function of time  shown in Fig.~1  of Paper~A is quite typical.  The
dependence on the size of the core is shown here in Fig.~\ref{chisdp of
rc} for several U20 models.  The ``core radius'' used is the empirical
one given by the radius at which the surface density of the mass
species used for the SDP comparison reaches half its central value.
Clearly, the minima in \chiSDP\ occur  before core bounce and then
becomes somewhat larger both as core collapse proceeds and subsequently
in the post-collapse phase.  The root of this difference lies in the
details of the matching between the data and models in the central part
of the cluster.  It is well to bear in mind that the core is not
resolved in the star count data, so the observational core radius is
not well defined.  For the same reason, the model core radius is not
constrained very well by the observations. The central  cusp is also
seen in the surface brightness profile but that is dominated by a few
bright stars.  \markcite{Lauzeral \et (1992)} manage to obtain a core
radius of 0.6 pc by removing the bright stars, but, as was shown in
\markcite{Drukier (1993)}, the small number of stars in the core still
precludes the exclusion of a smaller core radius.   \markcite{Meylan \&
Mayor (1991)} compiled all the surface brightness data onto one system
and measured a core radius of 0.22 pc by fitting the data to multi-mass
King models. This $r_c$ is consistent with those listed in
Table~\ref{outpar}. As discussed in \markcite{Drukier~\et (1993)}, the
SDP within 1 pc is somewhat flatter than beyond this radius, but it
never does flatten out.  In the models, as the core continues to
collapse, the region outside the core shows a single power-law slope
and the straight region around 0.5 pc occupies the region where the
observed SDP has a bump. Closer to core collapse the models fit the
central density better, but the match around 0.5 pc is much poorer. As
the cluster continues to evolve, the central slope continues to
flatten. This gives the increase in \chiSDP\ during the post-collapse
phase.

\section{Details of some good matches}
\label{models}

I will now turn to the presentation of some specific models which come
the closest to satisfying all the observational constraints.  None is perfect,
but the differences between them provide further information with
regard to the stellar content of NGC~6397. The models selected are from
the model sets described in Paper A. For the model sets showing
large variations in \chiSDP\ and \chiMF\ with the age of the model, I
have chosen the well-fitting model closest to the point of intersection
between the the two $\chi^2$ curves.  The one exception is model U20-C where
a somewhat older model has been used. For the U10, $r_t=18.5$ pc model
set, the one without a time dependence to the quality of the matches,
two models have been chosen, one at the young limit and one at the
old.  I have also included one model from the poorly fitting model set
X2 with $r_t=18.5$pc. Table~\ref{inpar} lists the initial parameters for each
of the nine models. Details about the state of each model at its best
matching time are listed in Table~\ref{outpar}. The comparisons
between some of the models and the observations are shown in Figs.~\ref{GM031}
to \ref{X018}.

Figure \ref{GM031} shows the match to the data of model
U20-A. The mass functions are all consistent with the general shapes of
the mass functions, although the fine details are missed by the
models.  This is as expected since the IMFs are simply combinations of
power laws and the true MFs appear to have more structure than can be
represented by this model.  The number of stars in each mass
function is also well fit for the du Pont:if and FRST mass functions;
in common with all the other models the number of stars in the du
Pont:out field is  underestimated.  The surface density profile is well
matched at all radii with the possible exception of the very center.
Other models, as will be seen, deviate by much larger amounts from the
observed central value, so in comparison model U20-A is quite
successful. The velocity dispersions are also well matched. The main
problem with this model is that it requires the turn-off mass to be
0.86\Msolar\ and hence the age to be 12 Gyr. As discussed in
Paper A, this problem is generic to any model set with too high a central
mean mass, ie. one with lots of heavy remnants.

Comparing Fig. \ref{GM031} (model U20-A, U20 with $r_t=17$ pc model
set) with Fig. \ref{GG000} (model U20-B, U20 with $r_t = 18.5$ pc model
set) Fig. \ref{GK079} (model U20-D, U20 with $r_t=21$ pc model set)
shows the effect of increasing the tidal radius for the U20 IMF.
Several systematic trends are visible in the comparison of these
models.  First, the  modeled du Pont:if mass function becomes steeper
with increasing $r_t$ for $m>0.4\Msolar$. This suggests that the
amount  of mass segregation within the half-mass radius decreases with
increasing $r_t$.  On the other hand, the shapes of the outer two mass
functions are much the same for all values of $r_t$.  The number of
stars in the modeled du Pont:out MF increases with increasing $r_t$, as
would be expected, and the quality of the  match to this mass function
similarly improves.  On the other hand, the central surface density
decreases with increasing tidal radius, as does the velocity
dispersion. The U20, $r_t = 20$ pc  model set model U20-C is consistent
with these trends. The trends in the central SDP, the du Pont:if MF and
the velocity dispersion are all consistent with a decrease in the total
mass in the inner part of the cluster with increasing tidal radius.
Considering the appropriate lines in Table~\ref{outpar}, the total mass
varies by about 6\% for the four models, but the mass within the inner
parsec decreases by 1/4 as the tidal radius increases.  This trend is
much less pronounced in models U10-B2 and U10-C, but the mass
difference and mass fraction in heavy remnants is also smaller with
these two models.  As is often the case in this modeling, post-facto
explanations are possible, but they are not easy to generalize in the
face of differences in the IMF.

Until here the models shown have all had the same IMF, U20.  The effects
of changing the upper mass limit, and hence the fraction of neutron
stars, is shown by comparing Fig.~\ref{GG000} with Fig.~\ref{R036} for
the U30 IMF and Figs.~\ref{T063} and \ref{T038} for the U10 IMF.  The
details of these IMFs are given in Table~\ref{IMFS} and more fully in
Table~2 of Paper~A. Unusually, model U10-B1 (Fig.~\ref{T063}) has been
caught deep in core collapse. This demonstrates that such a state is
not excluded, but is just usually not favored by the matching
procedure.  The age of the model is also compatible with the isochrone
age. The velocity data is equally consistent for both U10 models and
the differences in the mass function matches are not significant.

In a slightly different vein, Fig.~\ref{L105} shows the comparison for
model L05-B, a model with the same upper limit as the U20 IMF (20
\Msolar) but which extends down to 0.05 \Msolar.  This model shows some
larger deviations from the observations.  The central surface density
is much flatter in shape than observed and the velocity dispersion,
while within the observational errors, is systematically low.  The mass
function at the high mass end is somewhat steeper than observed.  All
three of these deviations indicate that this IMF has too few massive
stars relative to the number of low mass stars.  Model L05-B is, of all
the models shown, the largest compromise between the SDP and the MFs.
As was shown in Paper~A, the L05 model set showed the strongest time
dependence for both \chiMF\ and \chiSDP.  At the young extreme, there
is a model which gives a better fit to the SDP and at a much older
limit there are models which give better matches to the MFs. At both
extremes the deficiencies in the match to the other data are
accentuated, and in all the L05 models the velocity dispersions are
systematically low. This all suggests that the L05 IMF has too high a
proportion of low mass stars and that NGC~6397 contains only about 20\%
by mass of stars with masses less than 0.2\Msolar.  On the positive
side, the age of this model, 14 Gyr, is consistent with the isochrone
age of NGC~6397.

Unlike the model sets discussed until now, the X2 and NNS model sets
produced no models which adequately matched the observations. The
primary problem lay in the surface density profile as I show in
Fig.~\ref{X018}.  I show this model at the time of the minimum in
\chiMF\ at about the same time before core collapse as most of the
other models discussed here.  However, for the X2 and NNS model sets
the time of core collapse produces a maximum in \chiSDP. At this time
the SDP profile has a central logarithmic slope of about $-1.3$, much
steeper than the observed value of $-0.9$. This indicates that the
turn-off stars have about the same mass as the stars dominating the
core. If we consider the radius at which the enclosed mean mass equals
the mean mass of the main-sequence turn-off (as defined by the age of
the model), then for a typical X2 model this radius is about 0.2 pc
compared with 0.4 to 0.8 pc for the better matching model sets.
Interestingly, the best matching U10 model, U10-B1, has the smallest of
these radii, the X2 model set has just taken this trend too far.  The
model mass functions are somewhat too steep at the high mass end of
the mass functions. This suggest that a flatter IMF is required.  The
conclusion I draw from this is that NGC~6397 contains about
2500\Msolar\ in neutron stars. Model X2-B contains only
1400\Msolar\ and this is clearly not adequate, but model U10-B1, with
2600\Msolar\ is quite satisfactory.

\section{Conclusions}

These detailed comparisons lead to the following conclusions about
NGC~6397.
\begin{enumerate}
\item The total mass of the cluster is $6.6\pm 0.5\times 10^4 \Msolar$.
\item Approximately 2500 \Msolar\ of this is in neutron stars.
\item Given that the absolute integrated V magnitude of NGC~6397 is $M_V=-7.02$
(Djorgovski 1993) the global
mass-to-light ratio is $1.2 \pm 0.1 \left(\Msolar/L_\odot\right)_V$.
\item The mass function probably flattens for stars less massive than
the observed limit. The mass fraction in stars with masses less than
0.2 \Msolar\ is probably less than \onethird\ and is more like
\slantfrac{1}{5}.
\item The core radius is unresolved in these data, but is probably less
than 0.3 pc (11\arcsec).
\item The tidal radius of the cluster is $17\pm 4$ pc reflecting
a probable perigalactic distance of $2.5\pm 0.7$ kpc. These numbers are
somewhat uncertain as they come from the mass function fits and not
from the surface density profile. Crowding by field stars prevents direct
observation of the tidal cut off.
\end{enumerate}

There are a couple of systematic problems with these matches. One is
that the models usually underestimate the number of stars in the du
Pont:out mass function and the second is that the model velocity
dispersion is lower than the observed velocity dispersions.  The
mass-to-light ratio derived here is substantially lower than the
mass-to-light ratio of 2. which  \markcite{Meylan \& Mayor (1991)}
derived from fitting  King models.  The King models give  a  higher
mass, $10^5\Msolar$, for the cluster, and thus a higher $M/L$.  The fit
of \markcite{Meylan \& Mayor} required increasing anisotropy in the
outer part of the cluster and it could be that the deficiencies in the
models here also indicate a requirement for anisotropy.  An anisotropic
velocity tensor would result in a somewhat higher line-of-sight
velocity dispersion and in higher stellar densities at radii where
anisotropy is important when compared with an isotropic distribution.

\markcite{Weinberg (1994)} has run some similar models to those run here;
excluding stellar evolution, but including disk shocking.  At the
galactocentric radii of the models discussed here, continuous disk
shocking destroys all clusters with $W_0<6.5$ in less than a Hubble
time. While these are only preliminary results, they do suggest that
this is an important effect for clusters like NGC~6397.  The
requirement for a high initial concentration would further restrict the
range of initial parameters which lead to acceptable models. Somewhat
higher initial masses may also be possible. The strength of tidal shocking
will depend on the cluster's orbit, so it is difficult to extrapolate
from Weinberg's results.

In this binary paper I have demonstrated that the type of Fokker-Planck
model described here can produce models capable of matching a set of
detailed observations of a globular cluster. The caveats are that
anisotropy in the velocity dispersion may be required and that disk
shocking may have an important role to play.  In the continued absence
of an accurate procedure for including velocity anisotropy in
Fokker-Planck models (but see \markcite{Takahashi 1993}), the present
approach is still useful. A large investment of computer time is
required to test various possible IMFs, but once the general nature of
the relationship between the models and the observations is clear, a
more faster and more systematic study is possible. I have also shown
how to make the comparisons and how to extract specific information
from the ensemble of models. A study such as this one can give definite
values for the mass of a cluster and can serve as a good guide to the
relative abundances of both the heavy and light unobserved stars.
Adequate data sets would also give information on the radial structure
of the cluster including the core and limiting radii. Information on
the latter can be extracted from the mass functions even when field
star contamination prevents direct observation of the boundary.  Due to
the large number of parameters available for constructing the models,
as many constraints as possible are desirable. This first extensive
comparison between Fokker-Planck modeling and detailed observations of
a single cluster is quite encouraging both for confidence in the models
and for the ability to use the models to interpret the observations.

\acknowledgments

This work was supported by funding from NSERC of Canada and PPARC of the U.K.

\clearpage

\begin{table}
\caption{IMF definitions (with masses in \Msolar.)}
\begin{tabular}{lccl}
\hline
\hline
IMF & $m_{min}$ & $m_{max}$ & Comments\tablenotemark{a}\cr
\hline
U10 & 0.1 & 10. & $x =1.5, m<0.4$; $x=0.9, m>0.4$\cr
U20 & 0.1 & 20. & $x=1.5, m<0.4; x=0.9, m>0.4$\cr
U30 & 0.1 & 30. & $x=1.5, m<0.4; x=0.9, m>0.4$\cr
L05 & 0.05 & 20. & $x=1.5, m<0.4; x=0.9, m>0.4$\cr
X2 & 0.1 & 20. & $x=1.5, m<0.4; x=0.9, 0.4<m<2.; x=2, m>2.$\cr
NNS & 0.1 & 20. & as U20, but all stars with initial masses $>8$ are assumed to
escape\cr
\hline
\hline
\end{tabular}
\tablenotetext{a}{$x$ is the mass spectral index for a power-law mass function
of the form $dN\propto m^{-(1+x)} dm$.}
\label{IMFS}
\end{table}

\begin{table}
\caption{Preferred model masses}
\begin{tabular}{llcc}
\hline\hline
IMF & $r_t$ & $M$ at 16 Gyr & $M$ at best time\tablenotemark{a} \\
    & (pc)  & ($10^5 \Msolar$) & ($10^5 \Msolar$) \\
\hline
U30 & 18.5 & 0.73 & 0.68\\
U20 & 17.  & 0.71\tablenotemark{b} & 0.68\\
U20 & 18.5 & 0.70 & 0.66\\
U20 & 20.  & 0.70 & 0.65\\
U20 & 21.  & 0.69 & 0.66\\
L05 & 18.5 & 0.71 & 0.69\\
U10 & 18.5 & 0.63 & 0.63\\
U10 & 20.  & 0.63 & 0.62\\
X2  & 18.5 & 0.59\tablenotemark{c} & \\
NNS & 18.5 & 0.61\tablenotemark{c} &\\
\hline\hline
\end{tabular}
\tablenotetext{a}{time of minima in the mean $\chi^2$ (see
Table~\protect{\ref{outpar}} below and Fig.~7 of Paper~A.)}
\tablenotetext{b}{extrapolated}
\tablenotetext{c}{average over all models}
\label{optimal mass}
\end{table}

\begin{table}
\caption{Initial parameters of models}
\begin{tabular}{llcccccc}
\hline\hline
model & IMF & $r_t$ & $W_0$ & $M_0$  & \rlrt & $r_h$ & $t_rh$ \\
      &     &  (pc) &       &($10^5 \Msolar$) & & (pc) & (Gyr) \\
\hline
U30-B & U30 & 18.5  & 5.50  & 5.50 & 0.66  & 3.6   & 1.6    \\
U20-A & U20 & 17.   & 5.12  & 5.19 & 0.66  & 3.6   & 1.7    \\
U20-B & U20 & 18.5  & 6.36  & 5.79 & 1.00  & 4.5   & 2.5    \\
U20-C & U20 & 20.   & 5.12  & 4.59 & 0.66  & 4.0   & 1.9    \\
U20-D & U20 & 21.   & 6.69  & 5.19 & 1.10  & 5.0   & 2.8    \\
L05-B & L05 & 18.5  & 4.50  & 4.50 & 0.76  & 4.8   & 4.9    \\
U10-B1& U10 & 18.5  & 5.75  & 4.00 & 0.84  & 3.9   & 2.0    \\
U10-B2& U10 & 18.5  & 5.50  & 5.00 & 0.90  & 4.7   & 3.0    \\
U10-C & U10 & 20.   & 5.75  & 5.00 & 0.96  & 5.2   & 3.4    \\
X2-B  & X2  & 18.5  & 5.50  & 4.00 & 1.02  & 5.0   & 3.4    \\
\hline\hline
\end{tabular}
\label{inpar}
\end{table}

\begin{table}
\caption{Parameters of models at the best matching time}
\begin{tabular}{lccccccccccc}
\hline\hline
model & Age  & $M_{TO}$   & $r_{\bar m}$\tablenotemark{a}& \chiMF & \chiSDP &
Mass & $r_h$ & $r_c$ & $M\left(<1{\hbox{\rm pc}}\right)$ & NS mass &
$f_{<.2}$\tablenotemark{b} \\
      & (Gyr)& (\Msolar)     & (pc)               &        &         &($10^5
\Msolar$)& (pc) & (pc) & ($10^5 \Msolar$)& (\Msolar) \\
\hline
U30-B  & 12.2 &  0.86  & 0.79 & 1.4   & 1.9     & 0.70 & 3.6   & 0.29   & 0.104
& 9700 & 0.20\\
U20-A & 12.1 &  0.86  & 0.68 & 1.4   & 1.4     & 0.70 & 3.3   & 0.24   & 0.114
& 8600 & 0.19\\
U20-B & 12.0 &  0.86  & 0.69 & 1.4   & 1.5     & 0.68 & 3.6   & 0.26   & 0.100
& 7800 & 0.20\\
U20-C & 13.7 &  0.83  & 0.81 & 1.4   & 2.0     & 0.66 & 3.9   & 0.32   & 0.089
& 7600 & 0.21\\
U20-D & 12.2 &  0.86  & 0.69 & 1.6   & 1.7     & 0.67 & 4.1   & 0.33   & 0.083
& 6700 & 0.23\\
L05-B  & 13.8 &  0.83  & 0.56 & 1.9   & 1.9     & 0.72 & 4.1   & 0.42   & 0.081
& 6300 & 0.34\\
U10-B1  & 14.1 &  0.82  & 0.39 & 1.6   & 1.1     & 0.66 & 3.2   & 0.02   &
0.101 & 2600 & 0.20\\
U10-B2  & 16.2 &  0.79  & 0.43 & 1.6   & 1.1     & 0.64 & 3.2   & 0.35   &
0.098 & 3000 & 0.18\\
U10-C  & 17.8 &  0.77  & 0.52 & 1.4   & 1.1     & 0.63 & 3.5   & 0.25   & 0.093
& 2900 & 0.19\\
X2-B  & 18.1 &  0.77  & 0.23 & 2.2   & 2.7     & 0.58 & 2.8   & 0.14   & 0.112
& 1400 & 0.18\\
\hline\hline
\end{tabular}
\tablenotetext{a}{Radius at which the mean mass of the enclosed stars equals
the turn-off mass $M_{TO}$.}
\tablenotetext{b}{Mass fraction in stars with $m<0.2\Msolar$.}
\label{outpar}
\end{table}

\clearpage

\begin{figure}
\caption{Plot of the quality of fit to the mass functions (\chiMF) vs. total
mass for a representative sample of 21 U20 models.  The sharp
corners indicate the sampled times for each run.  Mass decreases with
time. Note the cluster of minima around $7\times 10^4\Msolar$.}
\label{mass vs. chi}
\end{figure}

\begin{figure}
\caption{Plot of minima in \chiMF\ for all the U20 models. Models with
large \chiMF\ are a poor match to the data on other grounds and can
be excluded from the overall mass estimate. The typical optimal mass
is $6.9\times 10^5\Msolar$ for the U20 IMF.}
\label{mass vs. chi 2}
\end{figure}

\begin{figure}
\caption{Optimal mass for all the models showing such minima plotted
against the model age for the eight model sets with well-fitting models
in Paper~A.  The arrangement of the model sets is the same as in Fig.~7
in Paper~A. The IMF of the model set is listed across the top and $r_t$
(in pc) is listed at the right. The U20, $r_t=18.5$ pc model set has the
best sampled parameter space and the thickness of the distribution
should reflect the uncertainty in the mass estimator. }
\label{mass grid}
\end{figure}

\begin{figure}
\caption{Plot of \chiSDP\ vs. the empirical core radius for the same
models as in Fig.~\protect{\ref{mass vs. chi}}. The models start at
top-right and evolve downwards and then to the left.  The minima in
$r_c$ are at core collapse.  Post collapse evolution takes them upwards
and to the right.  Note that the minima occur before core collapse and
that \chiSDP\ increases in the post collapse phase. }
\label{chisdp of rc}
\end{figure}

\begin{figure}
\caption{Comparison diagram for model U20-A in the U20, $r_t=17$
 pc model set.  Clockwise from upper left: The surface density profile;
 the FRST mass function; (top) the du Pont:if mass function and
 (bottom) the du Pont:out mass function; and the velocity dispersion
profile.  The dashed line in the mass function panels indicates the
shape of the IMF.
 }
\label{GM031}
\end{figure}

\begin{figure}
\caption{As Fig.~\protect{\ref{GM031}} for model U20-B in the
U20, $r_t=18.5$ pc model set.}
\label{GG000}
\end{figure}

\begin{figure}
\caption{As Fig.~\protect{\ref{GM031}} for model U20-D in the U20,
 $r_t=21$ pc model set.}
\label{GK079}
\end{figure}

\begin{figure}
\caption{As Fig.~\protect{\ref{GM031}} for model U30-B in the U30,
 $r_t=18.5$ pc model set.}
\label{R036}
\end{figure}

\begin{figure}
\caption{As Fig.~\protect{\ref{GM031}} for model U10-B1 in the U10,
 $r_t=18.5$ pc model set.}
\label{T063}
\end{figure}

\begin{figure}
\caption{As Fig.~\protect{\ref{GM031}} for model U10-B2 in the U10,
 $r_t=18.5$ pc model set.}
\label{T038}
\end{figure}

\begin{figure}
\caption{As Fig.~\protect{\ref{GM031}} for model L05-B in the L05,
 $r_t=18.5$ pc model set.}
\label{L105}
\end{figure}

\begin{figure}
\caption{As Fig.~\protect{\ref{GM031}} for model X2-B in the X2,
 $r_t=18.5$ pc model set.}
\label{X018}
\end{figure}
\clearpage
\end{document}